\documentclass[journal, draftclsnofoot, onecolumn, 11pt]{IEEEtran}
\usepackage{algorithm,algorithmic,amsbsy,amsmath,amssymb,epsfig,bbm,mathrsfs,fancyhdr,fancyvrb,subfigure,url,color}
\usepackage{amsmath}
\usepackage{amssymb}
\usepackage{mathtools}
\usepackage{enumerate}
\usepackage{paralist}
\usepackage{graphicx}
\usepackage{epstopdf}
\graphicspath{{Figures/}}
\usepackage{subfigure}
\usepackage{cite}
\usepackage{array}
\usepackage{booktabs}
\usepackage{comment}
\usepackage{cases}
\usepackage{algorithm}
\usepackage{algorithmic}

\usepackage{amssymb}
\usepackage{amsthm} 

\usepackage{etoolbox} 

\usepackage[mathscr]{euscript} 

\newcounter{myoptimizationproblemctr}
\newenvironment{myoptimizationproblem}{
   \bigskip\noindent
   \refstepcounter{myoptimizationproblemctr}
   $(\mathbf{P\themyoptimizationproblemctr})$ 
   }{}   

\newtheorem{theorem}{Theorem}

\newtheorem{proposition}[theorem]{Proposition}
\newtheorem{corollary}{Corollary}

\begin{document}

\title{Distributed Resource Allocation in D2D-Enabled  Multi-tier Cellular Networks: An Auction Approach}

%

\author{\IEEEauthorblockN{Monowar Hasan and Ekram Hossain} \\
\IEEEauthorblockA{Department of Electrical and Computer Engineering,  University of Manitoba, Canada \\
Email: $\lbrace$monowar\_hasan, Ekram.Hossain$\rbrace$@umanitoba.ca }
\and
\thanks{A part of this work appeared in proceedings of IEEE ICC 2015.}}


\maketitle

\begin{abstract}
Future wireless networks are expected to be highly heterogeneous with the co-existence of macrocells and small cells as well as provide support for device-to-device (D2D) communication. In such muti-tier heterogeneous systems centralized radio resource allocation and interference management schemes will not be scalable. In this work, we propose an auction-based distributed solution to allocate radio resources in a muti-tier heterogeneous network. We provide the bound of achievable data rate and show that the complexity of the proposed scheme is linear with number of transmitter nodes and the available resources. The signaling issues (e.g., information exchange over control channels) for the proposed distributed solution is also discussed. Numerical results show the effectiveness of proposed solution in comparison with a centralized resource allocation scheme.

\end{abstract}

\begin{IEEEkeywords}
Multi-tier cellular networks, device-to-device (D2D) communication, distributed resource allocation, auction method.
\end{IEEEkeywords}

\section{Introduction} \label{sec:intro}

 The future generation (i.e., 5G) of cellular wireless networks are expected to be a mixture of network tiers of different sizes, transmit powers, backhaul connections, different radio access technologies that are accessed by an unprecedented numbers of heterogeneous wireless devices \cite{toshiba_5g}. The multi-tier cellular architecture of 5G networks, where small cells (e.g., femto and pico cells) are underlaid on the macrocells and also there is provisioning for device-to-device (D2D) communications, is expected to not only increase the coverage and capacity of the cell but also improve the broadband user experience within the cell. Since uncoordinated power and
spectrum allocation for small cells as well as D2D user equipments (DUEs) can cause severe interference to the other receiving nodes,  efficient resource allocation and interference management is one of the fundamental research challenges for such multi-tier heterogeneous networks. Due to the nature of the resource allocation problem in multi-tier cellular networks, centralized solutions are computationally expensive and also incur huge signaling overhead. Therefore, distributed or semi-distributed solutions with low signaling overhead are desirable where the network nodes (such as small cell base stations [SBSs] and DUEs)  perform resource allocation independently or by the minimal assistance of macro base stations (MBSs).

In this paper, we use the concept of auction from Economics and present a \textit{distributed solution} to the resource allocation problem for a LTE-A based D2D-enabled multi-tier cellular network. The term {\em distributed}   refers to the fact that the SBSs and the DUEs independently determine the allocation with the minimal assistance of MBS.  The auction approach allows us to develop a polynomial time-complexity algorithm  which provides near-optimal performance. The auction process evolves with a bidding process, in which unassigned agents (e.g., transmitters) raise the cost and bid for the resources simultaneously. Once the bids from all the agents are available, the resources are assigned to the highest bidder.

Despite the fact that there are ongoing research efforts to address the resource allocation problems for two-tier networks (e.g., \cite{prabo_journal, distscn, auction_femto_vcg}) as well as for D2D communications (e.g., \cite{d2d_eefi, qos_d2d_4}), the distributed solutions for the radio resource allocation problems in a generic D2D-enabled multi-tier scenario has not been studied comprehensively in the literature. In \cite{prabo_journal}, a distributed solution is proposed for a two-tier network using evolutionary game. A utility-based distributed resource allocation scheme for small cell networks is proposed in \cite{distscn} using interference pricing. In \cite{d2d_eefi} and \cite{qos_d2d_4}, energy and QoS-aware resource allocation scheme is proposed for D2D communication.  Auction theory has also been used in the context of wireless resource allocation problems (e.g., \cite{auc_sbchnnel, auction_multihop, auction_nec, auction_femto_vcg}). An auction-based subchannel allocation for OFDMA and multihop systems is presented in \cite{auc_sbchnnel} and \cite{auction_multihop}, respectively. However, the power allocation is not considered. A resource allocation approach for multi-cell OFDMA system using the concept of auction is presented in \cite{auction_nec}. The uplink spectrum-sharing resource allocation problem in a sparsely deployed femtocell networks is considered in \cite{auction_femto_vcg} and the authors proposed an auction-based solution.  However, the works in \cite{d2d_eefi, qos_d2d_4, auc_sbchnnel, auction_multihop, auction_nec} consider single-tier systems and the works in \cite{prabo_journal, distscn, auction_femto_vcg} do not consider D2D communication. 


Different from the above works, we apply the auction method to solve the radio resource allocation problem in a heterogeneous muti-tier network.  We consider a multi-tier network consisting a MBS, a set of SBSs (such as pico and femto base stations) and corresponding small cell user equipments (SUEs), as well as DUEs. There is a common set of radio resources (e.g., resource blocks [RBs]) available to the network tiers (e.g., MBS, SBSs
and DUEs). The SUEs and DUEs use the available resources (e.g., RB and power level) in an underlay manner as long as the interference caused to the macro tier (e.g., macro user equipments [MUEs]) remains below a given threshold. The goal of resource allocation is to allocate the available RBs and transmit power levels to the SUEs and DUEs in order to maximize the spectral efficiency without causing significant interference to the MUEs. 

The main contribution of this work is a low-complexity decentralized solution to the radio resource allocation problem in a multi-tier cellular system. The key feature of the proposed approach is that it incurs polynomial time complexity and low overhead for information exchange. We analyze the complexity and the optimality of the solution. To this end, we also discuss the applicability of the proposed approach in a practical LTE-A based system. 

The rest of this paper is organized as follows. The system model, related assumptions, and the resource allocation problem is presented in Section \ref{sec:sys_mod}. The auction-based distributed solution is developed in Section \ref{sec:auc_dist_sol}. We present the numerical results in Section \ref{sec:numerical} before we conclude the paper in Section \ref{sec:conclusion}.

 \section{System Model and Problem Formulation} \label{sec:sys_mod}

\begin{figure}[!t]
\centering
\includegraphics[width = 3.0 in]{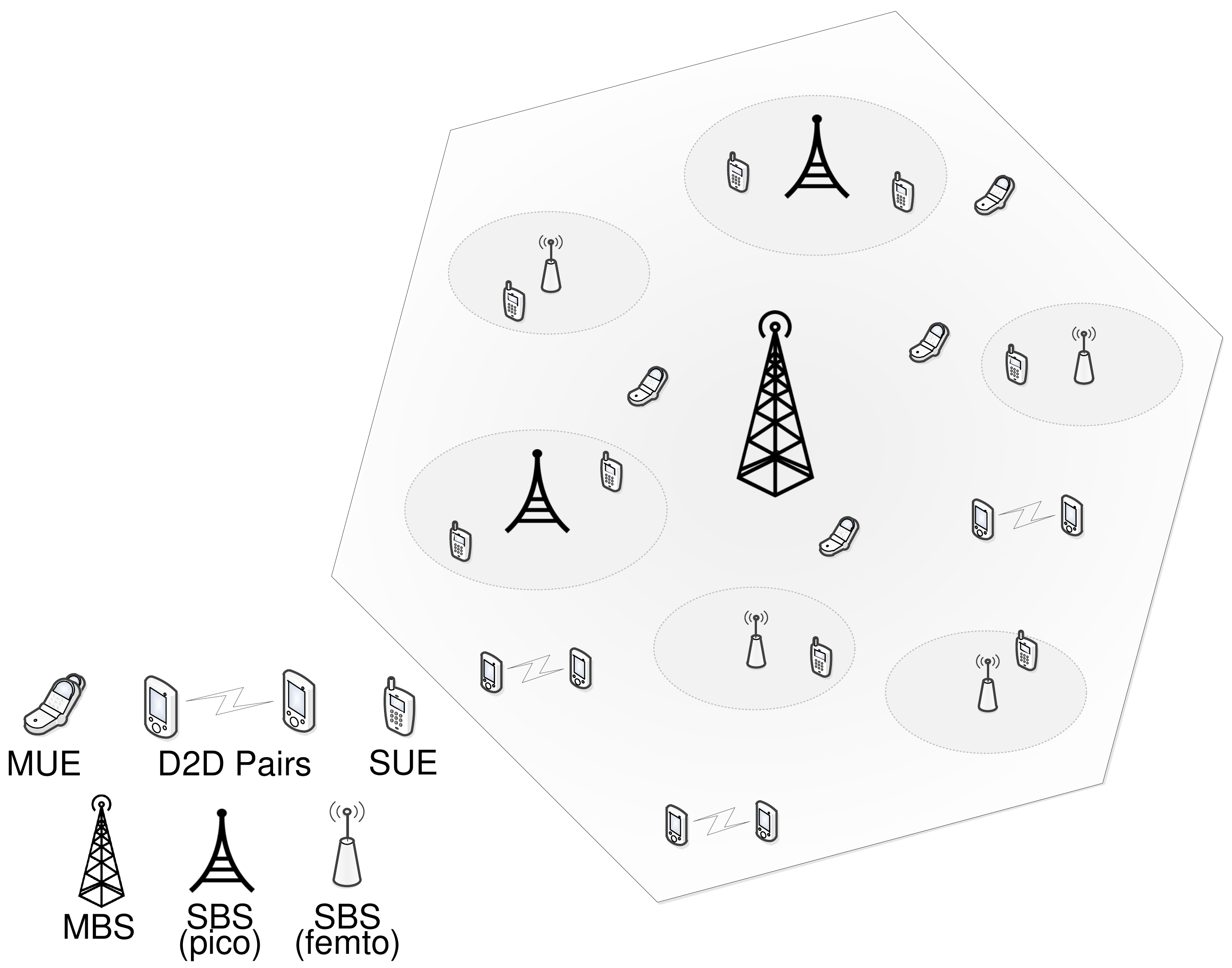}
\caption{A D2D-enabled mutil-tier cellular network. The small cells (such as pico and femto cells) as well as the D2D pairs are underlaid within the macrocell by reusing the same radio resources.} 
\label{fig:sys_mod}
\end{figure}

 \subsection{Network Model and Assumptions}
 
 Let us consider a transmission scenario in a D2D-enabled multi-tier network as shown in Fig. \ref{fig:sys_mod}. The network consists of one MBS and a set of $M$ cellular MUEs, i.e., $\mathcal{U}^{\mathrm m} = \lbrace 1,2,\cdots, M \rbrace$. There are also $D$ D2D pairs and $S$ closed-access SBSs (such as pico and femto base stations) located within the coverage area of the MBS. The set of SBSs is denoted by $\mathcal{S} = \lbrace 1, 2, \cdots S\rbrace$. We assume that each SBS serves only one SUE during a transmission interval\footnote{The scheduling of SUEs by the SBSs is not within the scope of this work.}. The set of active SUEs is given by $\mathcal{U}^{\mathrm s}$ where  $|\mathcal{U}^{\mathrm s}| =  S$. 
 The set of D2D pairs is denoted as $\mathcal{U}^{\mathrm d} = \lbrace 1,2,\cdots, D \rbrace$. In addition, the $d$-th element of the sets $\mathcal{U}^{\mathrm d_{T}}$ and $\mathcal{U}^{\mathrm d_{R}}$ denotes the transmitter and receiver UE of the D2D pair $d \in \mathcal{U}^{\mathrm d}$, respectively. The set of UEs in the network is given by $\mathcal{U} = \mathcal{U}^{\mathrm m} \cup \mathcal{U}^{\mathrm s} \cup \mathcal{U}^{\mathrm d}$. For ease of presentation, we refer to the SBSs (SUEs) and D2D transmitters (receivers) as underlay transmitters (receivers). Therefore, we denote $\mathcal{K}^{\mathrm T} = \mathcal{S} \cup \mathcal{U}^{\mathrm d_T}$ the set of underlay transmitters (e.g., SBSs and transmitting D2D UEs) and by $\mathcal{K}^{\mathrm R} =  \mathcal{U}^{\mathrm s} \cup \mathcal{U}^{\mathrm d_{R}}$ the set of underlay receivers (e.g., SUEs and receiving D2D UEs). Hence, $K = S + D$ denotes the total number of underlay transmitters/receivers. 

We refer to the small cells and D2D pairs as \textit{underlay tier}. The SBSs and DUEs are underlaid within the \textit{macro tier} (e.g., MBS and MUEs) since both the macro tier and the underlay tier (e.g., SBSs, SUEs and D2D pairs) use the same set $\mathcal{N} = \lbrace 1, 2, \cdots, N \rbrace$ of orthogonal RBs. In the considered multi-tier  network model, each of the network tiers (e.g., macro tier and underlay tier consisting with small cells and D2D UEs) has different transmit power, coverage region and specific set of users. 

Each transmitter node in the underlay tier (e.g., SBS and D2D transmitter) selects one RB from the available $N$ RBs. In addition, the underlay transmitters are capable of selecting the transmit power from a finite set of power levels, i.e., $\mathcal{L} = \lbrace 1, 2, \cdots, L \rbrace$.  
Each SBS and D2D transmitter should select a suitable RB-power level combination. This RB-power level combination is referred to as \textit{transmission alignment} \cite{prabo_journal}. For each RB $n \in \mathcal{N}$, there is a predefined threshold $I_{\mathrm{TH}}^{(n)}$ for maximum aggregated interference caused by the underlay tier to the macro tier. We assume that value of $I_{\mathrm{TH}}^{(n)}$ is known to the underlay transmitters by using the feedback control channels. An underlay transmitter is allowed to use the particular transmission alignment as long as the cross-tier interference to the MUEs is within the threshold limit.

We assume that the user association to the base stations (either MBS or SBSs) is completed prior to resource allocation. In addition, the potential DUEs are discovered during the D2D session setup by transmitting known synchronization or reference signals (i.e., beacons) \cite{network_asst_d2d}. According to our system model, only one MUE is served on each RB to avoid co-tier interference within the macro tier. However, multiple underlay UEs (e.g., SUEs and DUEs) can reuse the same RB to improve the spectrum utilization. This reuse causes severe cross-tier interference to the MUEs, and also co-tier interference within the underlay tier. Therefore, an efficient resource allocation scheme will be required. 
 
 \subsection{Achievable Data Rate}
 
 The MBS transmits to the MUEs using a fixed power $p_{\mathfrak{m}}^{(n)} > 0$ for $\forall n$.  For each underlay transmitter $k \in \mathcal{K}^{\mathrm T}$, the transmit power over the RBs is determined by the vector $\mathbf{P}_{\mathrm k} = \left[ p_k^{(1)}, p_k^{(2)}, \cdots, p_k^{(N)} \right]^{\mathsf{T}}$ where $p_k^{(n)} \geq 0$ denotes the transmit power level of transmitter $k$ over RB $n$. The transmit power $p_k^{(n)}, ~\forall n$  must be selected from the finite set of power levels $\mathcal{L}$. Note that if the RB $n$ is not allocated to transmitter $k$, the corresponding power variable $p_k^{(n)} = 0$. Since we assume that each underlay transmitter selects only one RB, only one element in the power vector $\mathbf{P}_{\mathrm k}$ is non-zero.
 
 
 We denote by $g_{i, j}^{(n)}$ the channel gain for the links $i$ and $j$ over RB $n$.  For the SUEs, we denote by $u_k$ the SUE associated to SBS $k \in \mathcal{S}$, and for the DUEs, $u_k$ refers to the receiving D2D UE of the D2D transmitter $k \in \mathcal{U}^{\mathrm d_{T}}$. The received signal-to-interference-plus-noise ratio (SINR) for the any arbitrary SUE or D2D receiver, i.e., $u_k \in \mathcal{K}^{\mathrm R}, k \in \mathcal{K}^{\mathrm T}$ over RB $n$ is given by
 \begin{equation} \label{eq:sinr_underlay}
 \gamma_{u_k}^{(n)} = \frac{g_{k, u_k}^{(n)}p_{k}^{(n)}}{g_{\mathfrak{m}, u_k}^{(n)}p_{\mathfrak{m}}^{(n)} + \sum\limits_{\substack{ k^\prime \in  \mathcal{K}^{\mathrm T}, k^\prime \neq k  }} g_{k^\prime, u_k}^{(n)} p_{k^\prime}^{(n)} + ~\sigma^2}
 \end{equation}
 where $g_{k,u_k}^{(n)}$ is the link gain between the SBS and SUE (e.g., $u_k \in \mathcal{U}^{\mathrm s},  k \in \mathcal{S}$) or the link gain between the D2D UEs (e.g., $u_k \in \mathcal{U}^{\mathrm d_{R}}, k \in \mathcal{U}^{\mathrm d_T}$), and $g_{\mathfrak{m}, u_k}^{(n)}$ is the interference gain between the MBS and the UE $u_k$. In the SINR expression (\ref{eq:sinr_underlay}),  $\sigma^2 = N_0 W_{\mathrm{RB}}$ where $W_{\mathrm {RB}}$ is the bandwidth corresponding to an RB and $N_0$ denotes the thermal noise. Given the SINR, the data rate of the UE $u_k$ over RB $n$ can be calculated as $R_{u_k}^{(n)} = W_{\mathrm {RB}} \log_2 \left(1 +  \gamma_{u_k}^{(n)} \right)$. 
 
\subsection{Formulation of the Resource Allocation Problem}

The objective of radio resource (i.e., RB and transmit power) allocation problem is to obtain the assignment of RB and power level (e.g., transmission alignment) for the underlay UEs (e.g., D2D UEs and SUEs) that maximizes the achievable sum data rate. The RB and power level allocation indicator for any underlay transmitter $k \in \mathcal{K}^{\mathrm T}$ is denoted by a binary decision variable $x_{k}^{(n, l)}$, where
\begin{equation}
x_{k}^{(n, l)}  = \begin{cases}
 1, \quad  \text{if the transmitter $k$ is trasnmitting over RB $n$ with power level $l$} \\
 0, \quad \text{otherwise.}
\end{cases}
\end{equation}
The achievable data rate of an underlay UE $u_k$ with the corresponding transmitter $k$ is written as 
\begin{equation} \label{eq:rate_ue}
R_{u_k} = \sum\limits_{n = 1}^{N} \sum\limits_{l = 1}^{L}  ~x_{k}^{(n,l)}   W_{\mathrm {RB}} \log_2 \left(1 +  \gamma_{u_k}^{(n)} \right). 
\end{equation}
The aggregated interference experienced on RB $n$ is given by 
\begin{equation} \label{eq:ref_ue}
I^{(n)} = \sum\limits_{k =1}^{K}\sum\limits_{l = 1}^{L}x_{k}^{(n, l)} g_{k,m_k^*}^{(n)} p_k^{(n)}
\end{equation}
where $m_k^* = \underset{m}{\operatorname{argmax}}~ g_{k,m}^{(n)}, ~\forall m \in \mathcal{U}^{\mathrm m}$. The concept of reference user \cite{ref_user} is adopted to calculate the aggregated interference $I^{(n)}$ using (\ref{eq:ref_ue}). For any RB $n$, the interference caused by the underlay transmitter $k$ is determined by the highest gains between the transmitter $k$ and MUEs, e.g., the MUE $m_k^*$ who is the mostly affected UE by the transmitter $k$. Satisfying the interference constraints considering the link gain corresponding to the reference user will also satisfy the interference constraints for other MUEs. As mentioned earlier, an underlay transmitter is allowed to use a particular transmission alignment only when it does not violate the interference threshold to the MUEs, i.e., $I^{(n)} < I_{\mathrm{TH}}^{(n)}, ~\forall n$.

Let $\mathbf{X} = \left[x_{1}^{(1, 1)}, \cdots, x_{1}^{(1, L)}, \cdots, x_{1}^{(N, L)}, \cdots, x_{K}^{(N, L)} \right]^{\mathsf{T}}$ denote the resource (e.g., transmission alignment) allocation vector. Mathematically, the resource allocation problem can be expressed as follows:

\begin{myoptimizationproblem} \label{opt:combopt}
\vspace*{-2.0em}
\begin{subequations}
\begin{align}
\hspace{3em} \underset{\mathbf{X}}{\operatorname{max}} ~ \sum_{\substack{k =1}}^{K}  \sum_{n = 1}^{N} \sum_{l = 1}^{L}  x_{k}^{(n,l)}    & W_{\mathrm {RB}}  \log_2\left( 1 + \gamma_{u_k}^{(n)} \right)  \nonumber \\
 \text{subject to}~~ \sum_{k =1}^{K}\sum_{l = 1}^{L}x_{k}^{(n, l)} g_{k,m_k^*}^{(n)} p_k^{(n)} &< I_{\mathrm{TH}}^{(n)}, ~~\forall n \label{eq:opt_intf}\\
\sum_{n = 1}^{N} \sum_{l = 1}^{L} x_{k}^{(n,l)} &= 1, \quad ~~~\forall k  \label{eq:opt_rbpw}\\
x_{k}^{(n,l)} &\in \lbrace 0, 1 \rbrace,~\forall k, n, l \label{eq:opt_bin}
\end{align}
\end{subequations}
\end{myoptimizationproblem}
where \vspace*{-0.3em} \begin{equation} \label{eq:sinr_formulation}
\gamma_{u_k}^{(n)} = \frac{g_{k, u_k}^{(n)}p_{k}^{(n)}}{ g_{\mathfrak{m}, u_k}^{(n)}p_{\mathfrak{m}}^{(n)} + \sum\limits_{\substack{ k^\prime \in \mathcal{K}^{\mathrm{T}},\\ k^\prime \neq k }}^{K} \sum\limits_{l^\prime = 1}^{L} x_{k^\prime}^{(n,l^\prime)} g_{k^\prime, u_k}^{(n)} p_{k^\prime}^{(n)} + ~\sigma^2}.
\end{equation} 

The objective of the resource allocation problem $\mathbf{P\ref{opt:combopt}}$ is to maximize the data rate of the SUEs and DUEs. With the constraint in (\ref{eq:opt_intf}), the aggregated interference caused to the MUEs by the underlay transmitters on each RB is limited by a predefined threshold. The constraint in (\ref{eq:opt_rbpw}) indicates that the number of RB selected by each underlay transmitter should be one and each  transmitter can only select one power level at each RB. The binary indicator variable for transmission alignment selection is represented by the constraint in (\ref{eq:opt_bin}). Note that the decision variable $x_{k}^{(n, l)} = 1$ implies that $p_k^{(n)} = l$.

\begin{corollary}
The resource allocation problem $\mathbf{P\ref{opt:combopt}}$ is a combinatorial non-convex non-linear optimization problem. The complexity to solve the above problem using exhaustive search is of $\mathcal{O}\left( \left(NL \right)^{K} \right)$ and the centralized solution is strongly NP-hard for the large values of $K, N, L$. 
\end{corollary}

Due to mathematical intractability of solving the above resource allocation problem, in the following we present a distributed solution using tools from the  auction theory. The distributed solution is developed under the assumption that the system is feasible, i.e., given the resources and parameters (e.g., size of the network, interference thresholds etc.), it is possible to obtain an allocation that satisfies all the constraints of the original optimization problem. 

\section{Distributed Solution Using Auction Method} \label{sec:auc_dist_sol}

The resource allocation using auction is based on the bidding procedure, where the agents (i.e., underlay transmitters) bid for the resources (e.g., RB and power level). The transmitters select the bid for the resources based on the costs (e.g., the interference caused to the MUEs) of using the resource. The desired assignment relies on the appropriate selection of the bids. The unassigned transmitters simultaneously raise the cost of using resource and bid for the resources. When the bids from all the transmitters are available, the resources are assigned to the highest bidder.

In an auction-based assignment model, every resource $j$ associated with a cost  $c_j$  and each agent $i$ can obtain benefit $B_{ij}$ from the resource $j$. The net value (e.g., utility) that an agent $i$ can obtain from resource $j$ is given by $B_{ij} - c_j$. For an equilibrium assignment, every agent $i$ should be assigned with resource $j^\prime$ such that the condition $B_{ij^\prime} - c_{j^\prime} \geq \underset{j}{\operatorname{max}} \left\lbrace B_{ij} - c_{j} \right\rbrace - \epsilon$ is satisfied for all the agents, where $\epsilon > 0$ indicates the parameter related to the minimum bid requirement \cite{auction_org}. In the following we utilize the concept of auction in order to obtain the distributed solution of the resource allocation problem. 

\subsection{Utility and Cost Function}

Let us introduce the parameter $\Gamma_{u_k}^{(n, l)} \triangleq {\gamma_{u_k}^{(n)}}_{ \!\! \vert p_k^{(n)} = l }$ that denotes the achievable SINR of the UE $u_k$ over RB $n$ using power level $l$ (e.g., $p_k^{(n)} = l$) where $\gamma_{u_k}^{(n)}$ is given by (\ref{eq:sinr_formulation}). We express the data rate as a function of SINR. In particular, let $\mathscr{R}\left(\Gamma_{u_k}^{(n, l)}\right) =  W_{\mathrm {RB}} \log_2 \left(1 +  \Gamma_{u_k}^{(n, l)} \right)$ denotes the achievable data rate for transmitter $k$ over RB $n$ using power level $l$. 

The utility of an underlay transmitter for a particular transmission alignment is determined by two factors, i.e., the achievable data rate for a given RB and power level combination, and an additional cost function that represents the aggregated interference caused to the MUEs on that RB. In particular, the utility of the underlay transmitter $k$ for a given RB $n$ and power level $l$ is given by 
\begin{equation} \label{eq:sm_utility}
\mathfrak{U}_{k}^{(n,l)} = \nu_1 \mathscr{R}\left(\Gamma_{u_k}^{(n, l)}\right)  -  \left[ \nu_2  \left( I^{(n)} - I_{\mathrm{TH}}^{(n)} \right) \right]^+ 
\end{equation}
where the operator $\left[ \cdot \right]^+ = \max \left\lbrace 0, \cdot \right\rbrace$ and $\nu_1$, $\nu_2$ are the biasing factors which can be selected based on which network tier (i.e., macro or underlay tier) should be given priority for resource allocation \cite{prabo_journal}. Note that the term $\nu_2 \left( I^{(n)} - I_{\mathrm{TH}}^{(n)} \right) $ in (\ref{eq:sm_utility}) represents the cost (e.g., interference caused by underlay transmitters to the MUE) of using the RB $n$. In particular, when the transmitter $k$ is transmitting with power level $l$, the cost of using RB $n$ can be represented by
\begin{align} \label{eq:auc_cost_ra}
& c_{k}^{(n,l)} =  \nu_2 \left( I^{(n)} - I_{\mathrm{TH}}^{(n)} \right) \nonumber \\
&= \nu_2 \bigg(  g_{k,m_k^*}^{(n)} l +   \sum\limits_{\substack{ k^\prime \in \mathcal{K}^{\mathrm{T}}, \\ k^\prime \neq k}}\sum\limits_{l^\prime = 1}^{L}x_{k^\prime}^{(n, l^\prime)} g_{k^\prime,m_{k^\prime}^*}^{(n)} p_{k^\prime}^{(n)} - I_{\mathrm{TH}}^{(n)} \bigg).
\end{align}
Let the parameter $C_{k}^{(n,l)} = \left[ c_{k}^{(n,l)} \right]^+$ and accordingly the cost $C_{k}^{(n,l)} = 0$ only if $I^{(n)} \leq I_{\mathrm{TH}}^{(n)}$.  Using the cost term we can represent (\ref{eq:sm_utility}) as $\mathfrak{U}_{k}^{(n,l)} = B_{k}^{(n,l)} - C_{k}^{(n,l)}$, where $B_{k}^{(n,l)} = \nu_1 \mathscr{R}\left(\Gamma_{u_k}^{(n, l)}\right)$ is the weighted data rate achieved by transmitter $k$ using resource $\lbrace n,l \rbrace$. 


\subsection{Update of Cost and Bidder Information}

Let $\mathfrak{b}_{k}^{( n,l)}$ denotes the local bidding information available to transmitter $k$ for the resource $\lbrace n,l \rbrace$.  
In the beginning of the auction procedure, at any time slot $t$, each underlay transmitter updates the cost as
\begin{equation} \label{eq:auc_cstupdate}
C_{k}^{(n,l)}(t) = \underset{k^\prime \in \mathcal{K}^{\mathrm T}, k^\prime \neq k}{\operatorname{max}} \left\lbrace C_{k}^{(n,l)}(t-1), C_{k^\prime}^{(n,l)}(t-1) \right\rbrace.
\end{equation}
In addition, the information of highest bidder of the resource $\lbrace n, l \rbrace$ is obtained by 
\begin{equation} \label{eq:auc_hbidder}
\mathfrak{b}_{k}^{( n,l)}(t) = \mathfrak{b}_{k^*}^{( n,l)}(t-1)
\end{equation}
where $k^* = \underset{k^\prime \in \mathcal{K}^{\mathrm T}, k^\prime \neq k}{\operatorname{argmax}} \left\lbrace C_{k}^{(n,l)}(t-1), C_{k^\prime}^{(n,l)}(t-1) \right\rbrace$. 
When the cost of $\lbrace n, l \rbrace$ is greater than previous iteration and the transmitter $k$ is not the highest bidder, the transmitter needs to select a new transmission alignment, say, $\lbrace \hat{n},\hat{l}\rbrace$. The transmitter also increases the cost of the new resource $\lbrace \hat{n},\hat{l}\rbrace$ as $C_{k}^{(\hat{n}, \hat{l})}(t) = C_{k}^{(\hat{n}, \hat{l})}(t-1) +\Delta_k(t-1)$, where $\Delta_k(t-1)$ is given by
\begin{align} \label{eq:auc_costupdate}
\Delta_k(t-1) = \underset{\lbrace n^\prime, l^\prime \rbrace \in \mathcal{N}\times\mathcal{L}}{\operatorname{max}} \mathfrak{U}_{k}^{(n^\prime,l^\prime)}(t-1) - 
 \underset{\substack{\lbrace n^\prime, l^\prime \rbrace \in \mathcal{N}\times\mathcal{L} \\ n^\prime \neq \hat{n}, l^\prime \neq \hat{l} }}{\operatorname{max}} \mathfrak{U}_{k}^{(n^\prime,l^\prime)}(t-1) + \epsilon.
\end{align}
The term $\underset{\lbrace n^\prime, l^\prime \rbrace \in \mathcal{N}\times\mathcal{L}}{\operatorname{max}} \mathfrak{U}_{k}^{(n^\prime,l^\prime)}(t-1) - \underset{\substack{\lbrace n^\prime, l^\prime \rbrace \in \mathcal{N}\times\mathcal{L} \\ n^\prime \neq \hat{n}, l^\prime \neq \hat{l} }}{\operatorname{max}} \mathfrak{U}_{k}^{(n^\prime,l^\prime)}(t-1)$ physically represents the difference between the maximum and the second to the maximum utility value. In the case when the transmitter $k$ does not prefer to be assigned with a new resource, the allocation from the previous iteration will remain unchanged, i.e., $\mathbf{x}_k(t) = \mathbf{x}_k(t-1)$, where $\mathbf{x}_k = \left[ x_{k}^{(n,l)} \right]_{\forall n,l}^{\mathsf T}$.

\begin{algorithm} [!t]
\AtBeginEnvironment{algorithmic}{\footnotesize} 
\caption{Auction method for any underlay transmitter $k$}
\label{alg:auc_loc}
\begin{algorithmic}[1]   
\renewcommand{\algorithmicrequire}{\textbf{Input:}}

\renewcommand{\algorithmicensure}{\textbf{Output:}} 
\renewcommand{\algorithmicforall}{\textbf{for each}}
\renewcommand{\algorithmiccomment}[1]{\textit{/* #1 */}}

\REQUIRE Parameters from previous iteration: an assignment $\mathbf{X}(t-1) = \left[ \mathbf{x}_1(t-1), \cdots \mathbf{x}_K(t-1)  \right]^{\mathsf{T}}$, aggregated interference $\mathbf{I}(t-1) = \left[I^{(n)}(t-1)\right]_{\forall n}^{\mathsf{T}}$, 
cost of using resources $\mathbf{C}(t-1) = \left[ C_{k}^{(n,l)}(t-1)\right]_{\forall k,n, l}^{\mathsf{T}}$ and the highest bidders of the resources $\mathfrak{B}(t-1) = \left[ \mathfrak{B}_k(t-1) \right]_{\forall k}^{\mathsf{T}}$ where $\mathfrak{B}_k(\cdot) = \left[\mathfrak{b}_{k}^{( n,l)}(\cdot) \right]_{\forall n, l}^{\mathsf{T}}$.

\ENSURE The allocation variable $\mathbf{x}_k(t) = \left[x_{k}^{(n,l)}\right]_{\forall n, l}^{\mathsf{T}}$, updated costs $\mathbf{C}_k(t) = \left[ C_{k}^{(n,l)}(t)\right]_{\forall n, l}^{\mathsf{T}}$, and bidding information $\mathfrak{B}_k(t) = \left[\mathfrak{b}_{k}^{( n,l)}(t) \right]_{\forall n, l}^{\mathsf{T}}$ at current iteration $t$ for the transmitter $k$.

\STATE Initialize $\mathbf{x}_k(t) := \mathbf{0}$.

\STATE For all the resources $\lbrace n, l\rbrace \in \mathcal{N}\times \mathcal{L}$  update the cost and highest bidder using (\ref{eq:auc_cstupdate}) and (\ref{eq:auc_hbidder}), respectively.


%
%

\STATE $\lbrace \widetilde{n}, \widetilde{l} \rbrace := $ Non-zero entry in $\mathbf{x}_k(t-1)$.  ~\COMMENT{\footnotesize Resource assigned in previous iteration} 



\IF{ $C_{k}^{(\widetilde{n}, \widetilde{l})} (t) \geq C_{k}^{(\widetilde{n}, \widetilde{l})} (t-1)  $ \AND $\mathfrak{b}_{k}^{(\widetilde{n}, \widetilde{l})}(t) \neq k $ }

\STATE $\lbrace \hat{n}, \hat{l} \rbrace := \underset{\lbrace n^\prime, l^\prime \rbrace \in \mathcal{N} \times \mathcal{L}}{\operatorname{argmax}} \mathfrak{U}_{k}^{(n^\prime,l^\prime)}(t)$.  


\STATE $\mathfrak{I}^{(\hat{n})} := g_{k,m_k^*}^{(\hat{n})} \hat{l} +   I^{(\hat{n})}$. ~\COMMENT{\footnotesize Estimate interference level}

\IF{ $\mathfrak{I}^{(\hat{n})} < I_{\mathrm{TH}}^{(\hat{n})}$ }

\STATE Set $x_{k}^{(\hat{n},\hat{l})} := 1$.

\STATE Update the highest bidder for the resource $\lbrace \hat{n}, \hat{l} \rbrace $ as $\mathfrak{b}_{k}^{(\hat{n}, \hat{l}) }(t) := k$.

\STATE Increase the cost $C_{k}^{(\hat{n}, \hat{l})}(t) = C_{k}^{(\hat{n}, \hat{l})}(t-1) +\Delta_k(t-1)$ where $\Delta_k(t-1)$ is given by (\ref{eq:auc_costupdate}).



\ELSE 
\STATE  $\mathbf{x}_k(t) := \mathbf{x}_k(t-1)$.  ~\COMMENT{\footnotesize Keep the assignment unchanged}

\ENDIF

\ELSE 
\STATE $\mathbf{x}_k(t) := \mathbf{x}_k(t-1)$.  ~\COMMENT{\footnotesize Keep the assignment unchanged}

\ENDIF


\end{algorithmic}
\end{algorithm}

\subsection{Algorithm for Resource Allocation}

We outline the auction-based resource allocation approach in \textbf{Algorithm \ref{alg:auc_alg}}. Each transmitter locally executes \textbf{Algorithm \ref{alg:auc_loc}} and obtains a temporary allocation. After execution of \textbf{Algorithm \ref{alg:auc_loc}}, each underlay transmitter $k$ reports to the MBS the local information, e.g., choices for the resources, $\mathbf{x}_k = \left[ x_{k}^{(n,l)} \right]_{\forall n,l}^{\mathsf{T}}$. Once the information (e.g., output parameters from \textbf{Algorithm \ref{alg:auc_loc}}) from all the transmitters are available to the MBS, the necessary parameters (e.g., input arguments required by \textbf{Algorithm \ref{alg:auc_loc}}) are calculated and broadcast by the MBS. \textbf{Algorithm \ref{alg:auc_loc}} repeats in an iterative manner until the allocation variable $\mathbf{X} = \left[\mathbf{x}_k \right]_{\forall k}^{\mathsf{T}} = \left[x_{1}^{(1, 1)}, \cdots, x_{1}^{(1, L)}, \cdots, x_{1}^{(N, L)}, \cdots, x_{K}^{(N, L)} \right]^{\mathsf{T}}$ for two successive iterations remains unchanged.

\begin{algorithm} [!t]
\AtBeginEnvironment{algorithmic}{\footnotesize} 
\caption{Auction-based resource allocation}
\label{alg:auc_alg}
\begin{algorithmic}[1]   
\renewcommand{\algorithmicrequire}{\textbf{Input:}}
\renewcommand{\algorithmicensure}{\textbf{Output:}}
\renewcommand{\algorithmicforall}{\textbf{for each}}
\renewcommand{\algorithmiccomment}[1]{\textit{/* #1 */}}

\renewcommand{\algorithmicensure}{\textbf{Phase I: Initialization}}
\ENSURE

\STATE Estimate the channel state information from the previous time slot.

\STATE Each underlay transmitter $k \in \mathcal{K}^{\mathrm T}$ randomly selects a transmission alignment and reports to the MBS.

\STATE MBS broadcasts the assignment of all transmitters, aggregated interference of each RB, the costs and the highest bidders using pilot signals.

\STATE Initialize number of iterations $t := 1$.

\renewcommand{\algorithmicensure}{\textbf{Phase II: Update}}
\vspace*{0.5em}
\ENSURE

\WHILE{$\mathbf{X}(t) \neq \mathbf{X}(t-1)$ \AND $t < T_{\mathrm{max}}$}

\STATE Each underlay transmitter $k \in \mathcal{K}^{\mathrm T}$ locally runs the \textbf{Algorithm \ref{alg:auc_loc}} and reports the assignment $\mathbf{x}_k(t)$, the cost $\mathbf{C}_k(t)$ and the bidding information $\mathfrak{B}_k(t)$ to the MBS. 

\STATE MBS calculates the aggregated interference vector $\mathbf{I}(t)$; forms the vectors $\mathbf{X}(t)$,  $\mathfrak{B}(t)$, and $\mathbf{C}(t)$; and broadcast to the underlay transmitters.

\STATE Update $t := t+1$.
\ENDWHILE

\renewcommand{\algorithmicensure}{\textbf{Phase III: Transmission}}
\vspace*{0.5em}
\ENSURE

\STATE Use the resources (e.g., the RB and power levels) allocated in the final stage of update phase for data transmission.

\end{algorithmic}
\end{algorithm}

\subsection{Optimality, Complexity, and Signaling over Control Channels}

\begin{proposition} \label{prop:kepsilon}
The sum data rate obtained by the distributed auction algorithm is within $K \epsilon$ of the optimal solution.
\end{proposition}
\begin{IEEEproof}
Refer to \textbf{Appendix \ref{appsec:kepsilonproof}}.
\end{IEEEproof}

Since the proposed scheme satisfies the allocation constraints (\ref{eq:opt_rbpw})-(\ref{eq:opt_bin}), and also maintains the interference threshold given by (\ref{eq:opt_intf}), the solution obtained by the auction algorithm gives a lower bound of original resource allocation problem  $\mathbf{P\ref{opt:combopt}}$. As shown in the following proposition, when the allocation remain unchanged for at least $T \geq 2$ consecutive iterations, the complexity of the proposed solution is linear with number of underlay transmitters and the available resources.

\begin{proposition} \label{prop:timcom}
The auction algorithm converges to a fixed allocation with the number of iterations of $\mathcal{O}\left( T KNL \Upsilon \right)$ where $ \Upsilon = \left \lceil {\frac{\underset{k, n, l}{\operatorname{max}} B_{k}^{(n,l)} - \underset{k, n, l}{\operatorname{min}}  B_{k}^{(n,l)}}{\epsilon} } \right\rceil$.
\end{proposition}
\begin{IEEEproof}
Refer to \textbf{Appendix \ref{appsec:timcom}}.
\end{IEEEproof}

In the following, we discuss the applicability of the proposed method in a practical system. In the proposed solution, the MBS only needs the allocation vector $\mathbf{X}$ and the channel gain vector $\mathbf{G} = \left[ g_{k,m_{k}^*}^{(n)} \right]_{\forall k, n}^{\mathsf{T}}$ to calculate the aggregated interference vector, e.g.,  $\mathbf{I} = \left[I^{(n)}\right]_{\forall n}^{\mathsf{T}}$. Also note that at each iteration, the MBS and the transmitters need to exchange the allocation vector $\mathbf{X}$, information about highest bidders $\mathfrak{B}$, the cost vector $\mathbf{C}$ to the underlay transmitters. However, in LTE-A based systems, these information can easily be incorporated into the standard control channel messages. For example, the MBS can broadcast these information vectors using physical downlink control channel (PDCCH). Likewise, each of the underlay transmitters $k$ can inform the MBS the local assignment preferences $\mathbf{x}_k$, updated costs $\mathbf{C}_k$ and bidding information $\mathfrak{B}_k$ using physical hybrid-ARQ indicator channel (PHICH).

\section{Numerical Results} \label{sec:numerical}

\subsection{Simulation Setup}

We develop a MATLAB-based simulator and observe the performance of our proposed approach using simulations. We simulate a $300 ~\text{m} \times 300 ~\text{m}$ area where the MBS is located in the center of the cell and $M = 6$ MUEs are uniformly distributed within the cell radius. The SBSs and SUEs are uniformly distributed within the macro cell and small cell radius, respectively. The DUEs are located according to the clustered distribution model \cite{d2d_cluster_dist_model}. We obtain the minimum bid increment parameter $\epsilon$ by trial and error and set it to $100$. We choose number of RBs $N = 6$, biasing factors $\nu_1 = \nu_2 = 1$, and assume the interference threshold to be $-70$ dBm for all the RBs.

For modeling the  propagation channel, we consider distance-dependent path-loss and shadow fading; furthermore, the channel is assumed to  experience Rayleigh fading. In particular, we use realistic 3GPP propagation environment presented in \cite{3gpp_plmodel}. For example, the following path-loss equation is used to estimate path-loss between SBSs and SUEs as well as to the MUEs and DUEs: $\psi_{\mathrm{S}}(\ell)_{[dB]} = 38.46 + 20 \log(\ell) + \xi_{ss} + 10 \log(\varsigma)$. The path-loss between the MBS and MUEs as well as DUEs and SUEs is estimated as $\psi_{\mathrm{M}}(\ell)_{[dB]} = 15.3 + 40 \log(\ell) + \xi_{sm} + 10 \log(\varsigma) + \xi_{w}$. Similarly, the direct link gain between DUEs is given by $\psi_{\mathrm{D}}(\ell)_{[dB]} =  148+40 \log(0.001 \ell) + \xi_{sd} + 10 \log(\varsigma) + \xi_{w}$. In the path-loss equations, $\ell$ is the distance between nodes in meter, $\xi_{ss}$,  $\xi_{sm}$, $\xi_{sd}$ account for shadow fading and are modeled as a log-normal random variables, $\varsigma$ is an exponentially distributed random variable which represents the Rayleigh fading channel power gain, and $\xi_{w}$ denotes outdoor wall loss. Numerical results are averaged over different independent spatial network realizations and channel gains. The key simulation parameters and assumptions are listed in Table \ref{tab:sim_param}.

\begin{table}[!t]
\renewcommand{\arraystretch}{1.3}
\caption{Simulation Parameters}
\label{tab:sim_param}
\centering
\begin{tabular}{l|l}
\hline
\bfseries Parameter & \bfseries Values\\
\hline\hline
Cell layout & Isolated, hexagonal grid \\
Transmission bandwidth  & $1.08$ MHz \\
Macro and small cell radius & $300$ m and $30$ m \\
Distance between D2D peers & $15$ meter\\
Number of SUE per SBS & $1$ \\
Transmit power of MBS & $43$ dBm \\
Standard deviation for shadowing: & \\
\hspace{1em} Macrocell and D2D pairs  &  $8$ dB \\
\hspace{1em} Small cells & $4$ dB \\
Outdoor wall loss & 30 dB \\
Noise power spectral density & $-174$ dBm/Hz \\
\hline
\end{tabular}
\end{table}

\subsection{Results}

\begin{figure}[h t b]
\centering
\subfigure[]{\includegraphics[width = 4.0 in]{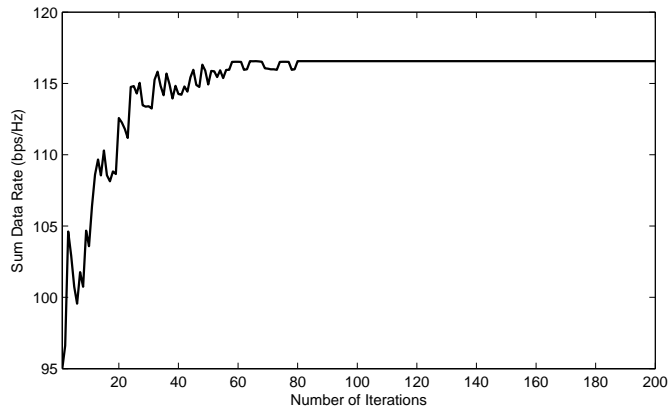}%
\label{fig:cn_sn}} 
\hfil 
\subfigure[]{\includegraphics[width = 4.0 in]{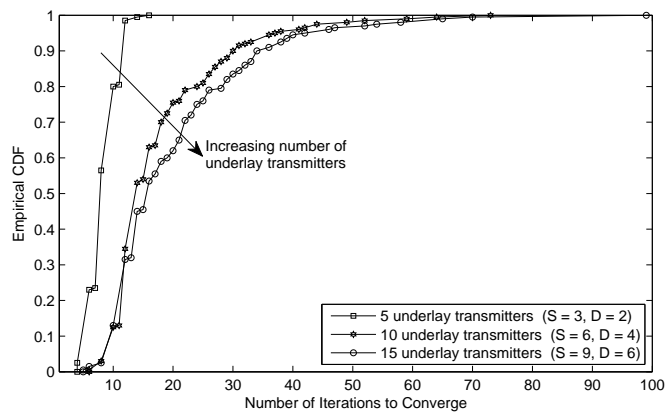}%
\label{fig:cn_cdf}}
\caption{(a) The convergence of proposed solution (for $L = 3$, e.g., $\mathcal{L} = \lbrace 3, 5, 7\rbrace$ dBm, $S = 6$, $D=4$), and (b) Empirical CDF of the number of iterations required for convergence using similar power levels as those for Fig. \ref{fig:cn_sn}.}
\label{fig:convergence}
\end{figure}

The convergence behavior of the proposed solution is depicted in Fig. \ref{fig:convergence}. We can find that the data rate of the network becomes stable within $100$ iterations. In order to observe the convergence behavior in different network density, we vary the number of underlay nodes (e.g., SBSs, SUEs and DUEs) and plot the empirical cumulative distribution function (CDF) of the number of  iterations required for convergence in Fig \ref{fig:cn_cdf}. The empirical CDF is defined as $\widehat{F}_\tau(\jmath) = \frac{1}{\tau} \displaystyle \sum_{i=1}^\tau \mathbb{I}_{[\zeta_i \leq \jmath]}$, where $\tau$ is the total number of simulation observations, ${\zeta}_i$ is the number of iterations required for convergence at the $i$-th simulation observation, and $\jmath$  represents the $x$-axis values in Fig. \ref{fig:cn_cdf}. The indicator function $\mathbb{I}_{[\cdot]}$ outputs $1$ if the condition $[\cdot]$ is satisfied and $0$ otherwise. As mentioned in \textbf{Proposition \ref{prop:timcom}}, the convergence to the fixed allocation depends on number of underlay transmitters. When the number of network nodes increases, the number of iterations required for convergence increases. This is because, the underlay transmitters need to execute the algorithm more times in order to obtain the updated bidding and cost information, and hence to find the fixed allocation. However, even in a moderately dense network with $30$ underlay nodes (e.g., $S + D = 9 + 6 = 15)$, the solution converges to a fixed rate within $100$ iterations.

\begin{figure}[!t]
\centering
\includegraphics[width = 4.0 in]{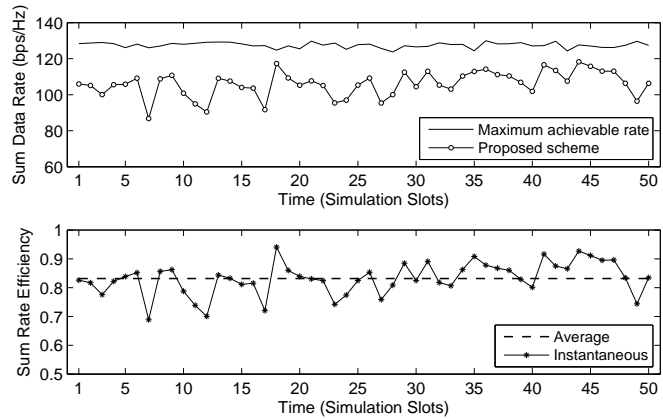}
\caption{Achievable sum rate of the proposed scheme and the upper bound. We consider $S = 3$, $D = 2$, and $\mathcal{L} = \lbrace 3, 5 \rbrace$ dBm.} 
\label{fig:comp_opt}
\end{figure}

To evaluate the performance of the proposed algorithm, in Fig. \ref{fig:comp_opt} we compare the the data rate with the upper bound obtained from the solution of the original optimization problem $\mathbf{P\ref{opt:combopt}}$. We obtain the upper bound (e.g., maximum achievable data rate) using exhaustive search. We examine the sum rate of the network in a period of $50$ time slots and average the results over different spatial realizations. We measure the sum rate efficiency as $\eta = \frac{R_{\mathrm{prop}}}{R_{\mathrm{max}}}$ where $R_{\mathrm{prop}}$ and $R_{\mathrm{max}}$ denote the sum data rate obtained from proposed scheme and maximum achievable sum rate, respectively. From this figure and as mentioned in \textbf{Proposition \ref{prop:kepsilon}}, it can be observed that the data rate of the proposed solution is within $K\epsilon$ of the maximum achievable data rate. Recall that the original resource allocation problem is an NP-hard problem and the computational complexity of exhaustive search to achieve the upper bound is of $\mathcal{O}\left( \left(NL \right)^{K} \right)$. The proposed scheme provides data rates which are close to $80 \%$ of the maximum achievable data rate, however, with significantly less computational complexity and signaling overhead.

\section{Conclusion} \label{sec:conclusion}

We have presented a polynomial time-complexity distributed solution approach for the future heterogeneous multi-tier cellular mobile communication systems. We analyze the  optimality and complexity the solution, and discuss about the applicability of the proposed scheme in a practical system. Numerical results have shown that the proposed solution performs close the upper bound of achievable data rate with significantly less complexity and minimal involvement of the central controller node. As an extension of this work, capturing the dynamics of misbehaving and selfish nodes in future dense networks using the other auction-based method (e.g., truthful auction) could be an interesting area of research.


\appendix
\numberwithin{equation}{section} 
\setcounter{equation}{0}  

\subsection{Proof of Proposition \ref{prop:kepsilon}} \label{appsec:kepsilonproof}

We construct the proof by using an approach similar to that presented in \cite{auction_nwflow}. The data rate obtained by any arbitrary assignment using auction method will satisfy the following condition: 
\begin{equation} \label{eq:auc_prof_ineqal}
\sum_{k=1}^{K} \nu_1 R_{u_k} \leq \!\!\!\!\!\! \sum_{\lbrace n, l \rbrace \in \mathcal{N} \times \mathcal{L}} \!\!\!\! \widehat{C}^{(n,l)} + \sum_{k=1}^{K} \underset{ \lbrace n, l \rbrace \in \mathcal{N} \times \mathcal{L}}{\operatorname{max}} \big\lbrace B_{k}^{(n,l)} -  \widehat{C}^{(n,l)} \big\rbrace 
\end{equation}
where $\widehat{C}^{(n,l)} = \underset{k^\prime \in \mathcal{K}^{\mathrm T}}{\operatorname{max}} C_{k^\prime}^{(n,l)} $  and $R_{u_k}$ is given by (\ref{eq:rate_ue}). 

The inequality in (\ref{eq:auc_prof_ineqal}) is satisfied since $$\sum\limits_{\lbrace n, l \rbrace \in \mathcal{N} \times \mathcal{L}} \widehat{C}^{(n,l)} = \sum\limits_{k=1}^{K} \sum\limits_{n=1}^{N} \sum\limits_{l=1}^{L} x_{k}^{(n,l)} C_{k}^{(n,l)}$$ and $$\sum\limits_{k=1}^{K} \underset{ \lbrace n, l \rbrace \in \mathcal{N} \times \mathcal{L}}{\operatorname{max}} \big\lbrace B_{k}^{(n,l)} -  \widehat{C}^{(n,l)} \big\rbrace \geq \sum\limits_{k=1}^{K} \sum\limits_{n=1}^{N} \sum\limits_{l=1}^{L} x_{k}^{(n,l)}\big(  B_{k}^{(n,l)} - \widehat{C}^{(n,l)}  \big).$$ Let the variable $A^* \triangleq \underset{\mathbf{X}}{\operatorname{max}} \sum\limits_{k=1}^{K} \nu_1 R_{u_k} = \sum\limits_{k=1}^{K} \sum\limits_{n = 1}^{N} \sum\limits_{l = 1}^{L} \nu_1 {x_{k}^{(n,l)}}  W_{\mathrm {RB}} \log_2 \left(1 +  \gamma_{u_k}^{(n)} \right)$ denotes the maximum achievable weighted sum rate.   In addition, let the variable $D^*$ be defined as
\begin{align}
D^* \triangleq \underset{\substack{\widehat{C}^{(n,l)} \\ \lbrace n, l \rbrace \in \mathcal{N} \times \mathcal{L}}}{\operatorname{min}} \Big\lbrace \sum_{\lbrace n, l \rbrace \in \mathcal{N} \times \mathcal{L}} \widehat{C}^{(n,l)} ~ + 
\sum_{k=1}^{K} \underset{ \lbrace n, l \rbrace \in \mathcal{N} \times \mathcal{L}}{\operatorname{max}} \big\lbrace B_{k}^{(n,l)} -  \widehat{C}^{(n,l)} \big\rbrace  \Big\rbrace.
\end{align}
Hence by definition we can write $A^* \leq D^*$. Since the final assignment and the set of costs are at equilibrium, for any underlay transmitter $k$, the condition $$\sum\limits_{n=1}^{N} \sum\limits_{l=1}^{L} x_{k}^{(n,l)}\big(  B_{k}^{(n,l)} - \widehat{C}^{(n,l)}  \big) \geq \underset{ \lbrace n, l \rbrace \in \mathcal{N} \times \mathcal{L}}{\operatorname{max}} \big\lbrace B_{k}^{(n,l)} -  \widehat{C}^{(n,l)} \big\rbrace - \epsilon$$ will hold. Consequently, we can obtain the following inequality:
\begin{align}
D^* &\leq \sum_{k=1}^{K} \Big( \sum_{n=1}^{N} \sum_{l=1}^{L} x_{k}^{(n,l)} \widehat{C}^{(n,l)} + \!\! \underset{ \lbrace n, l \rbrace \in \mathcal{N} \times \mathcal{L}}{\operatorname{max}} \big\lbrace B_{k}^{(n,l)} -  \widehat{C}^{(n,l)} \big\rbrace  \Big) \nonumber \\
&\leq \sum_{k=1}^{K}\sum_{n=1}^{N} \sum_{l=1}^{L} x_{k}^{(n,l)}  B_{k}^{(n,l)} + K \epsilon \nonumber \\
& \leq \sum_{k=1}^{K} \nu_1 R_{u_k} + K \epsilon \leq A^* + K \epsilon. 
\end{align}
Since $A^* \leq D^*$, the data rate achieved by the auction algorithm is within $K \epsilon$ of the optimal data rate $A^*$ and the proof follows. 

\subsection{Proof of Proposition \ref{prop:timcom}} \label{appsec:timcom}

From \cite{auction_nwflow} it can be shown that, in the worst case the total number of iterations in which a resource (e.g., transmission alignment) can receive a bid is no more than $ \Upsilon = \left \lceil {\frac{\underset{k, n, l}{\operatorname{max}} B_{k}^{(n,l)} - \underset{k, n, l}{\operatorname{min}}  B_{k}^{(n,l)}}{\epsilon} } \right\rceil$. Since each bid requires $\mathcal{O}\left(NL \right)$ iterations, and each iteration involves a bid by a single transmitter, the total number of iterations in \textbf{Algorithm \ref{alg:auc_alg}} is of $\mathcal{O}\left(  KNL \Upsilon \right)$. For convergence, the allocation variable $\mathbf{X}$ needs to be unchanged for at least $T \geq 2$ consecutive iterations. Hence, the overall running time of the algorithm is of $\mathcal{O}\left( T KNL \Upsilon \right)$. It is worth noting that for any transmitter node $k \in \mathcal{K}^{\mathrm T}$, the complexity of the auction process given by \textbf{Algorithm \ref{alg:auc_loc}} is linear with the number of resources for each of the iterations.

\bibliographystyle{IEEEtran}

\end{document}